\documentclass[10pt]{iopart}

\usepackage{graphics,epsfig,subfig}
\usepackage{amssymb,amsfonts,bm}
\usepackage{setstack}
\usepackage{mathrsfs}
\usepackage{color,psfrag}
\usepackage{iopams}
\usepackage{stackrel}

\usepackage{graphics,epsfig,subfig}
\usepackage{amssymb,amsfonts,bm}
\usepackage{mathrsfs}

\newcommand{\nn}{\nonumber}
\newcommand{\bb}{\begin{eqnarray}}
\newcommand{\ee}{\end{eqnarray}}

\renewcommand{\eref}[1]{(\ref{#1})}

\renewcommand{\e}{e}

\newcommand{\ff}{\frac{1}{2}}

\newcommand{\drm}{{\rm d}}
\renewcommand{\fl}{\hspace{-1cm}}
\newcommand{\com}{\textcolor{black}}

\newcommand{\ar}{a_{{\bm r}}}
\newcommand{\abr}{\bar a_{{\bm r}}}

\newcommand{\cg}[1]{c_{\bm{#1}}}
\newcommand{\cbg}[1]{\bar c_{\bm{#1}}}

\newcommand{\Br}{{\bm{r}}}

\newcommand{\Bk}{{\bm{k}}}

\newcommand{\ak}{a_{\bm{k}}}
\newcommand{\abk}{\bar a_{{\bm k}}}

\newcommand{\ccr}[1]{c_{\bm{#1}}}
\newcommand{\cdr}[1]{c^{\dagger}_{\bm{#1}}}

\newcommand{\Bd}{\bm{\delta}}
\newcommand{\trans}{\top}

\newcommand{\FF}{{\mathscr{F}}}

\newcommand{\fdfrac}[2]{\mbox{\footnotesize$\displaystyle\frac{#1}{#2}$}}

\newcommand{\tc}{\color{black}}

\newcommand{\Sint}{S_{\mathit{F+int}}}
\newcommand{\Seff}{S_{\mathit{eff}}}
\newcommand{\ZF}{Z_{\mathit{F}}}
\newcommand{\QF}{Q_{\mathit{F}}}
\newcommand{\SF}{S_{\mathit{F}}}
\newcommand{\AF}{A_{\mathit{F}}}

\begin{document}
\title{Itinerant electrons on dilute frustrated Ising lattices}
\author{Jean-Yves Fortin, Pierrick Lample}
\address
{\com{Laboratoire de Physique et Chimie Th\'eoriques,
CNRS UMR 7019, 
\\ Universit\'e de Lorraine,
F-54000 Nancy, France}
}
\ead{jean-yves.fortin@univ-lorraine.fr}
\begin{abstract}{We consider itinerant spinless fermions as moving defects in a dilute two-dimensional frustrated Ising system where they occupy site vacancies. Fermions interact via 
local spin fluctuations and we analyze coupled self-consistent mean-field equations of the Green functions after expressing the spin and fermion operators in terms of Grassmann variables. The specific heat and effective mass are analyzed with the solutions satisfying the symmetry imposed by the coupling layout. At low temperature, we find that these solutions induce stripes along the lines of couplings with the same sign, and that a low fermion density yields a small effective mass.}
\end{abstract} 

\section{Introduction}
%
Cooperative phenomena in correlated electronic systems yield to the emergence of spatial structures of the charge densities due to environmental constraints. For example, hole segregation was evidenced in the superconductor La$_{1.6-x}$Nd$_{0.4}$Sr$_x$CuO$_4$ and interpreted as the presence of a stripe modulation \cite{Cheong:1994,Tranquada:1995} which is assumed to be responsible for the suppression of superconductivity. One important question is how to probe for the possible existence of 
coherent states leading to superconductivity without the help of phonons \cite{Monthoux:2007}, possibly  using the indirect coupling induced by magnetic fluctuations.
In this paper we focus on the physical properties that emerge between itinerant fermions in a dilute frustrated spin environment. The indirect coupling between fermions is mediated by spin fluctuations via site dilution where fermions replace the missing spins, mimicking hole doping in two-dimensional
cuprate materials, but with classical spin fluctuations. The itinerancy leads to spin disorder however it is favorable for the fermions to move along the boundary of spin domains in order to lower the energy. In a different context, similar models such as the migration of electrons on ionic bonds have been studied theoretically \cite{scesney:70,sikakana:91}, also with classical moving defects \cite{selke:02}, and itinerant electrons coupled to a spin environment are known to generate various domain structures \cite{cencarikova:10,Yin:2010}.
Models of coupled lattices between spins and fermions were introduced to investigate metal-insulator interfaces \cite{Mondaini:2014} with the help of quantum Monte Carlo methods. In a previous paper \cite{Fortin:2021}, we analyzed a similar system of itinerant fermions occupying the empty sites of a
two-dimensional dilute Ising model. The dilution is known to suppress the magnetic order, but
spin fluctuations induce an indirect correlation between electrons. For this kind of model, we
observed the divergence of the effective mass in the critical region, and a diminution of its value
at low temperature, whereas the fermions behave like slightly heavy quasiparticles in the high temperature region. In the present paper, we would like to investigate the properties of these
itinerant fermions in the presence of frustration which does not display criticality but instead a macroscopic ground state degeneracy.

The paper is organized as follow. In section \ref{sect_model}, we introduce the quantum Hamiltonian
describing the interaction between fermions that occupy the vacancies of a classical spin model displaying frustration at low temperature with a highly degenerate ground state. The method used in the present paper is based on Grassmann techniques \cite{Clusel:2009} for the spin and fermion sectors \cite{lichtenstein:17}. The resulting action can be partially reduced by integrating over
the spin components and self-consistent equations for the Green functions and self-energies are analyzed under the mean-field approximation on an effective action, see section \ref{sect_coupling}. A homogeneous solution is given in section \ref{sect_homog}, and an investigation of a stripe solution satisfying the symmetry of the spin coupling configurations is developed in subsection \ref{sect_stripes}. Section \ref{sect_num} is dedicated to the Monte-Carlo study of the classical equivalent of the quantum model, with fermions replaced by itinerant classical particles obeying Markov dynamics, and finally a discussion in given in the conclusions, section \ref{sect_conc}.
%
%
\section{Model Hamiltonian for the frustrated dilute spin system \label{sect_model}}
\begin{figure}[!htb]
\centering
\subfloat[Piled up dominoes model]{
\includegraphics[angle=0,scale=0.4,clip]{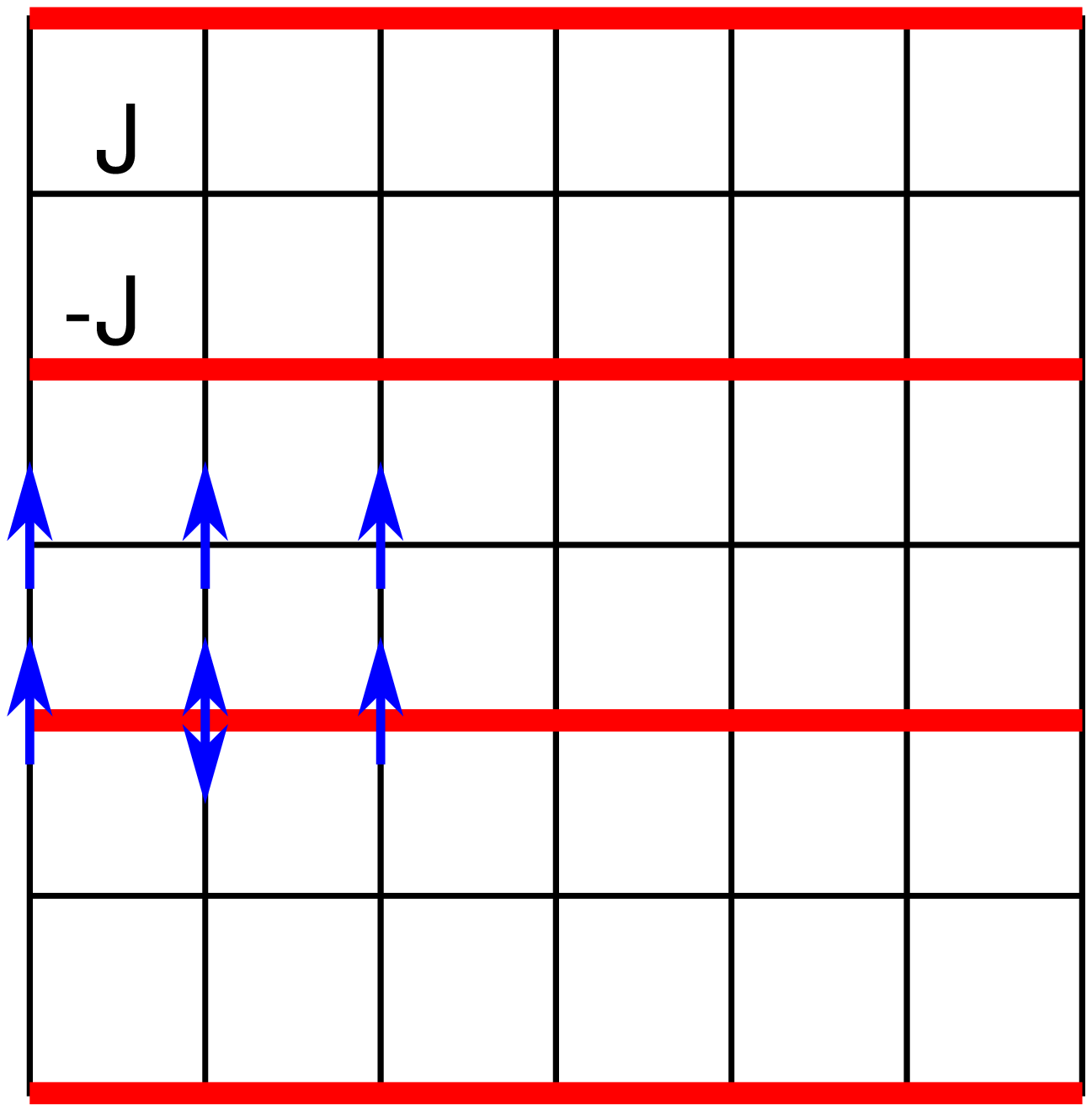}
}
\qquad
\subfloat[Zig-zag model]{
\includegraphics[angle=0,scale=0.4,clip]{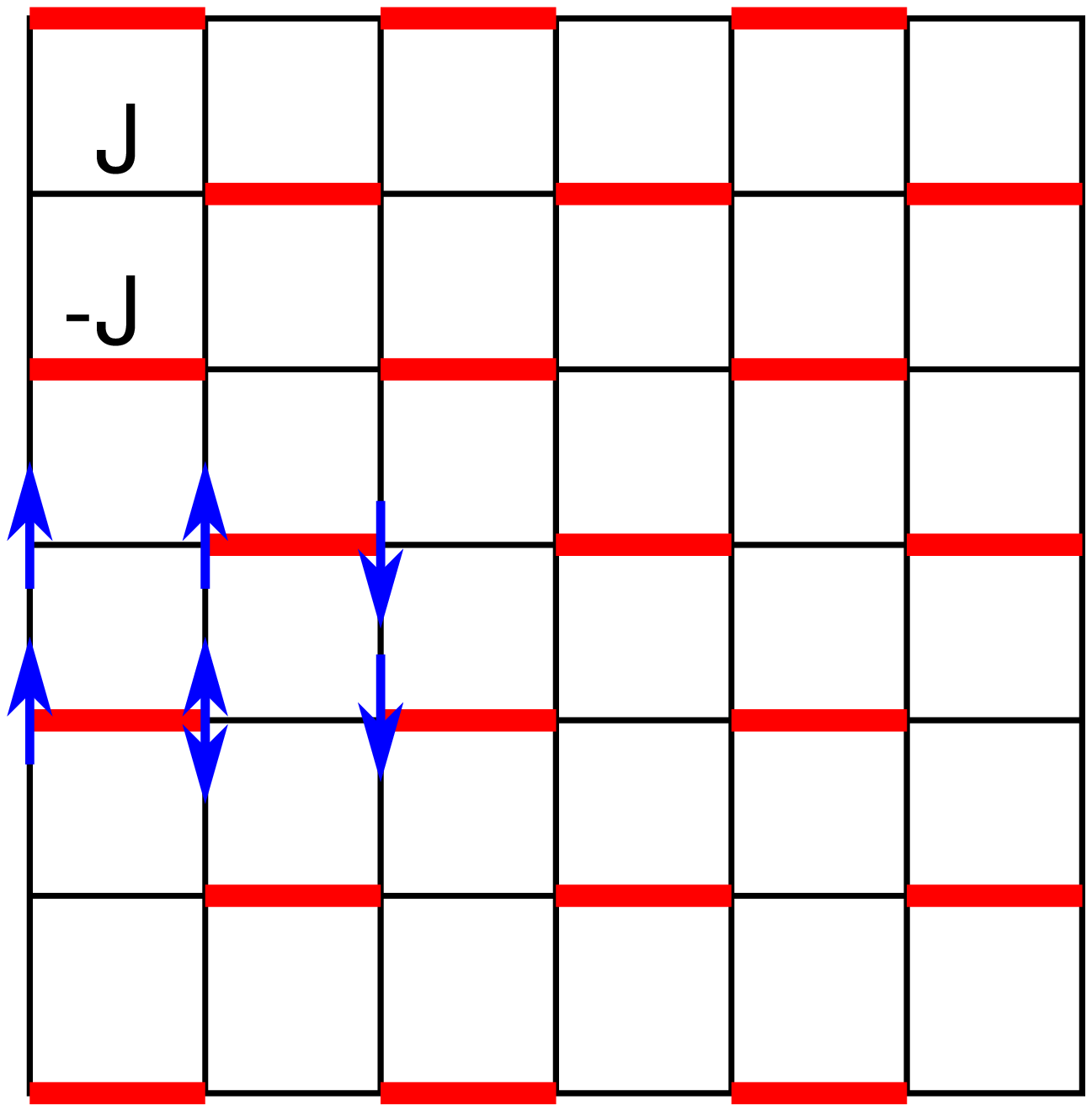}
}
\caption{\label{fig1}
Two frustrated lattice configurations with periodic and antiferromagnetic couplings $-J$ (red) and
ferromagnetic ones $J$ (black). Case (a) corresponds to the piled up dominoes model (PUD) , and the second to the zig-zag dominoes model (ZZD), see \cite{Andre:1979}. Each plaquette is frustrated. The double arrow indicates a ground state spin with two allowed orientations.}
\end{figure}

%
In this section we introduce the main Hamiltonian of a system composed of Ising spins 
$\sigma_{\Br}=\pm 1$ and spinless fermions on a frustrated lattice of size 
$N=L\times L$. Fermions are described by operators $(\ccr{r},\cdr{r})$ and fermion numbers $n_{\Br}=\cdr{r}\ccr{r}$. When a fermion occupies a site $\Br=(m,n)$, the corresponding spin becomes inactive. Also, every fermion moves to a neighboring site with a kinetic or transfer energy $t$, which will be set to unity in the numerical applications. The Hamiltonian is then defined by
\bb\label{Ham}
H=-\frac{J}{2}\sum_{\Br,\Bd=\bm{\pm x,\pm y}}\epsilon_{\Br}\sigma_{\Br}\sigma_{\Br+\Bd}(1-n_{\Br})(1-n_{\Br+\Bd})-t\sum_{\Br,\Bd=\bm{\pm x,\pm y}}\cdr{r+\Bd}\ccr{r},
\ee
where $\epsilon_{{\Br}}=\pm 1$ is the sign function of the local couplings $J\epsilon_{\Br}$ between
spins, and $\Bd$ the discrete unit vectors of the lattice. Couplings are antiferromagnetic if $\epsilon_{\Br}=-1$, and ferromagnetic otherwise. We consider two simple models of Ising lattices in which the geometry of the couplings leads to spin frustration. By definition, a system is frustrated if the product of the four couplings around each plaquette is negative. This can be achieved by considering two simple configurations of couplings. 
The first one corresponds to the piled up dominoes model (PUD), and the second one to the zig-zag dominoes model (ZZD), as represented in figure \ref{fig1}. These models were previously extensively studied and exact solutions were given in the non dilute case \cite{Andre:1979} using transfer matrix methods. Among the principal results is the macroscopically degeneracy of the ground state and a non-zero entropy per site at zero temperature with a non-trivial value. There is also no phase transition at non zero temperature for the case where all the couplings have the same absolute value, and the system remains paramagnetic.
The local couplings are defined for these two models by the function $\epsilon_{\Br}=\epsilon_{m,n}$ such that 
\bb\nn
\epsilon_{m,n}=(-1)^{n}\;{\mathrm{for\,the\,PUD\,model}},
\\ \label{def_eps}
\epsilon_{m,n}=(-1)^{m+n}\;{\mathrm{for\,the\,ZZD\,model}}.
\ee
The non-frustrated Ising case where $\epsilon_{m,n}=1$ was studied in a previous reference \cite{Fortin:2021} within a mean-field approximation on the Green functions of the fermionic sector.
%
%

We first consider only the spin Hamiltonian for the dilute case without the kinetic part of equation \eref{Ham}, with a coupling configuration displayed in figure \ref{fig1}(a) or (b)
\bb\label{HF}
H_{F}=-J\sum_{m,n}\Big (\epsilon_{m,n}y_{m,n}\sigma_{m,n}y_{m+1,n}\sigma_{m+1,n}+
y_{m,n}\sigma_{m,n}y_{m,n+1}\sigma_{m,n+1}\Big ),
\ee
where $y_{m,n}=1-n_{m,n}=0$ when a defect (hole or fermion) is present and $y_{m,n}=1$ otherwise when the site is occupied by a spin. 
The total Hamiltonian including kinetic part of the fermions, see equation \eref{Ham}, will be studied in section \ref{sect_coupling} below and an effective action will be derived. We should notice that
the spin part of the Hamiltonian \eref{HF} which contains the number operators $n_{\Br}$ does not commute with the kinetic part of the fermions in general.
Introducing the partition function $\ZF=\sum_{\sigma_{m,n}=\pm 1}e^{-\beta H_{F}}$, we can rewrite
the Boltzmann weights as
\bb\fl\nn
\ZF=\sum_{\sigma_{m,n}=\pm 1}\cosh(K)^{2N}
\prod_{m,n}(1+u\epsilon_{m,n}y_{m,n}\sigma_{m,n}y_{m+1,n}\sigma_{m+1,n})
(1+uy_{m,n}\sigma_{m,n}y_{m,n+1}\sigma_{m,n+1}),
\ee
with parameters $K=\beta J$ and $u=\tanh(K)$. The general case without vacancies can be solved using Grassmann variables and was derived extensively by Plechko \cite{plechko:85a,plechko:85b,plechko:97,plechko:99} in the case of various lattice geometries, and the partition function can be expressed as the Pfaffian of a quadratic form. In presence of vacancies, the Hamiltonian can be mapped onto an action with a quadratic and quartic parts in Grassmann variables representing the interaction contribution \cite{plechko:98,plechko:99p,plechko:10}, and therefore the partition function can not be expressed by a simple Pfaffian. However we can show using perturbative methods that the quartic part leads only to a small contribution \footnote{{\tc unpublished results, work to be submitted}}, see also \cite{Clusel:2009}.
In order to derive an effective action for the dilute frustrated models \eref{def_eps}, we follow in detail the method given by Plechko in a previous reference \cite{plechko:98} and introduce two pairs of Grassmann variables $(a_{m,n},\bar a_{m,n})$ and $(b_{m,n},\bar b_{m,n})$ such that
\bb\nn\fl
\ZF=\sum_{\sigma_{m,n}=\pm 1}\cosh(K)^{2N}
\prod_{m,n}\int d\bar a_{m,n}da_{m,n}e^{a_{m,n}\bar a_{m,n}}
(1+a_{m,n}y_{m,n}\sigma_{m,n})(1+u\epsilon_{m,n}\bar a_{m,n}y_{m+1,n}\sigma_{m+1,n})
\\ \fl\nn
\times\int d\bar b_{m,n}db_{m,n}e^{b_{m,n}\bar b_{m,n}}
(1+b_{m,n}y_{m,n}\sigma_{m,n})(1+u\bar b_{m,n}y_{m,n+1}\sigma_{m,n+1})
\\ \fl
=\sum_{\sigma_{m,n}=\pm 1}\cosh(K)^{2N}
\prod_{m,n}\int d\bar b_{m,n}db_{m,n}d\bar a_{m,n}da_{m,n}e^{a_{m,n}\bar a_{m,n}+b_{m,n}\bar b_{m,n}}
(A_{m,n}\bar A_{m+1,n})(B_{m,n}\bar B_{m,n+1}),
\ee
where the link operators in both directions are given by
\bb\nn
A_{m,n}=1+a_{m,n}y_{m,n}\sigma_{m,n},\;\bar A_{m,n}=1+u\epsilon_{m-1,n}\bar a_{m-1,n}y_{m,n}\sigma_{m,n}
\\ \label{AB}
B_{m,n}=1+b_{m,n}y_{m,n}\sigma_{m,n},\;\bar B_{m,n}=1+u\bar b_{m,n-1}y_{m,n}\sigma_{m,n}.
\ee
Moving the horizontal and vertical Grassmann factors inside the general product using mirror ordering principle, for which the computational details are not presented here, we obtain the partition function as a series of ordered products
\bb\fl\nn
\ZF=\cosh(K)^{2N}
\prod_{m,n}\int d\bar b_{m,n}db_{m,n}d\bar a_{m,n}da_{m,n}e^{a_{m,n}\bar a_{m,n}+b_{m,n}\bar b_{m,n}}
\\ \times
\sum_{\sigma_{m,n}=\pm 1}
\stackrel{\stackrel{n}{\longrightarrow}}{\prod_{n}}\Big [ \stackrel{\stackrel{m}{\longrightarrow}}{\prod_{m}}
\bar A_{m,n}\bar B_{m,n}A_{m,n} \stackrel{\stackrel{m}{\longleftarrow}}{\prod_{m}} B_{m,n} \Big ]
\ee
where the arrows indicate the order of the product, right or left. This allows for the summation over each individual spin variable using for example the fact that only even powers of $\sigma_{m,n}$ do not vanish: $\sum_{\sigma_{m,n}=\pm 1}\sigma_{m,n}^{2}=2$. We can then perform the exponentiation of the quadratic Grassmann factors
\bb\fl
\ZF=2^N\cosh(K)^{2N}
\prod_{m,n}\int d\bar b_{m,n}db_{m,n}d\bar a_{m,n}da_{m,n}\exp
\left (a_{m,n}\bar a_{m,n}+b_{m,n}\bar b_{m,n}\right )
\\ \nn\fl
\times\exp\left \{ y_{m,n}\left [a_{m,n}b_{m,n}+u(\epsilon_{m-1,n}\bar a_{m-1,n}+\bar b_{m,n-1})(a_{m,n}+b_{m,n})+u^2\epsilon_{m-1,n}\bar a_{m-1,n}\bar b_{m,n-1}\right ]
\right \}.
\ee
We then integrate over the pairs $(a_{m,n},b_{m,n})$ in order to reduce the number of variables, using 
the general identity for any linear Grassmann factors $L_{m,n}$ and $\bar L_{m,n}$
\bb
\int db_{m,n}da_{m,n}e^{y_{m,n}a_{m,n}b_{m,n}+a_{m,n}L_{m,n}+b_{m,n}\bar L_{m,n}}=y_{m,n}+\bar L_{m,n}L_{m,n}.
\ee
This yields the following integral over the remaining $2N$ Grassmann variables $(\bar a_{m,n},
\bar b_{m,n})$ which can be relabeled using the substitutions $\bar a_{m,n}\rightarrow -a_{m,n}$ and
$\bar b_{m,n}\rightarrow \bar a_{m,n}$
%
%
\bb
\nn
\ZF=2^N\cosh(K)^{2N}
\prod_{m,n}\int d\bar a_{m,n}da_{m,n}
\exp\left (-u^2y_{m,n}\epsilon_{m-1,n}a_{m-1,n}\bar a_{m,n-1}
\right )
\\ \nn
\times \prod_{m,n}\Big [y_{m,n}+a_{m,n}\bar a_{m,n}+uy_{m,n}(a_{m,n}+\bar a_{m,n})(\epsilon_{m-1,n}a_{m-1,n}-\bar a_{m,n-1})\Big ].
\ee
When a vacancy is present $y_{m,n}=0$, we have the simple result $\int d\bar a_{m,n}da_{m,n}a_{m,n}\bar a_{m,n}=1$, and when $y_{m,n}=1$ we obtain
\bb\nn\fl
\int d\bar a_{m,n}da_{m,n}\exp\left (-u^2\epsilon_{m-1,n}a_{m-1,n}\bar a_{m,n-1}
+a_{m,n}\bar a_{m,n}+
u(a_{m,n}+\bar a_{m,n})(\epsilon_{m-1,n}a_{m-1,n}-\bar a_{m,n-1})
\right )
\\
\fl
=\int d\bar a_{m,n}da_{m,n}e^{{\SF}_{m,n}}.
\ee
The total action $\SF=\sum_{m,n}{\SF}_{m,n}$ corresponding to the undiluted spin system can be written as
\bb\label{SF}
\SF
=\sum_{m,n}\Big (
a_{m,n}\bar a_{m,n}+
u(a_{m,n}+\bar a_{m,n})(\epsilon_{m-1,n}a_{m-1,n}-\bar a_{m,n-1})
-u^2\epsilon_{m-1,n}a_{m-1,n}\bar a_{m,n-1}\Big ).
\ee
Therefore we can write the partition function as
\bb
\label{Z_int}
\ZF=2^N\cosh(K)^{2N}
\prod_{m,n}\int d\bar a_{m,n}da_{m,n}e^{{\SF}_{m,n}}\left [(1-y_{m,n})a_{m,n}\bar a_{m,n}e^{-{\SF}_{m,n}}+y_{m,n}\right ],
\ee
where
\bb\label{Int}
a_{m,n}\bar a_{m,n}e^{-{\SF}_{m,n}}=a_{m,n}\bar a_{m,n}+u^2\epsilon_{m-1,n}a_{m,n}\bar a_{m,n}a_{m-1,n}\bar a_{m,n-1}.
\ee
Because the variables $y_{m,n}$ can take only values 0 or 1, the expression \eref{Z_int} can not
in principle be exponentiated. Indeed, we first need to factorize the terms $y_{m,n}$ in the brackets
in order to exponentiate the quadratic and quartic terms exactly, following \eref{Int}
\bb\nn
y_{m,n}\left (1+\frac{1-y_{m,n}}{y_{m,n}}a_{m,n}\bar a_{m,n}e^{-{\SF}_{m,n}}\right )
=y_{m,n}\exp\left (\frac{1-y_{m,n}}{y_{m,n}}a_{m,n}\bar a_{m,n}e^{-{\SF}_{m,n}}\right ).
\ee
This presents in principle a singularity in the exponential term at $y_{m,n}=0$.
But this expression is formally true since, after integrating over the Grassmann variables in the overall integral \eref{Z_int}, the result is a polynom in $y_{m,n}$ and does not contain any singularity when $y_{m,n}=0$. However, we can assume that the $y_{m,n}$ variables can be extended to real variables that take any positive value less than unity, as we will show in the density approximation presented in the section \ref{sect_coupling}, and we obtain
\bb\nn
\ZF=2^N\cosh(K)^{2N}
\prod_{m,n}\int y_{m,n}d\bar a_{m,n}da_{m,n}
\\ \label{S_int}
\times\exp\left ({\SF}_{m,n}+v_{m,n}
a_{m,n}\bar a_{m,n}+v_{m,n}u^2\epsilon_{m-1,n}a_{m,n}\bar a_{m,n}a_{m-1,n}\bar a_{m,n-1}
\right ),
\ee
where $v_{m,n}=(1-y_{m,n})/y_{m,n}$. The variables $v_{m,n}$ can be viewed as local coupling
parameters between the non-dilute model and the vacancies whose contribution is represented
by a quadratic and quartic forms. In general, we assume that the contribution of the quartic form in \eref{S_int} is small, since it contains in the continuous limit a double space derivative whose amplitude should be small for low momenta.

%
\subsection{Partition function for the non-dilute case \label{sect_PF}}
\begin{figure}[!htb]
\centering
\subfloat[Piled up dominoes model]{
\includegraphics[angle=0,scale=0.5,clip]{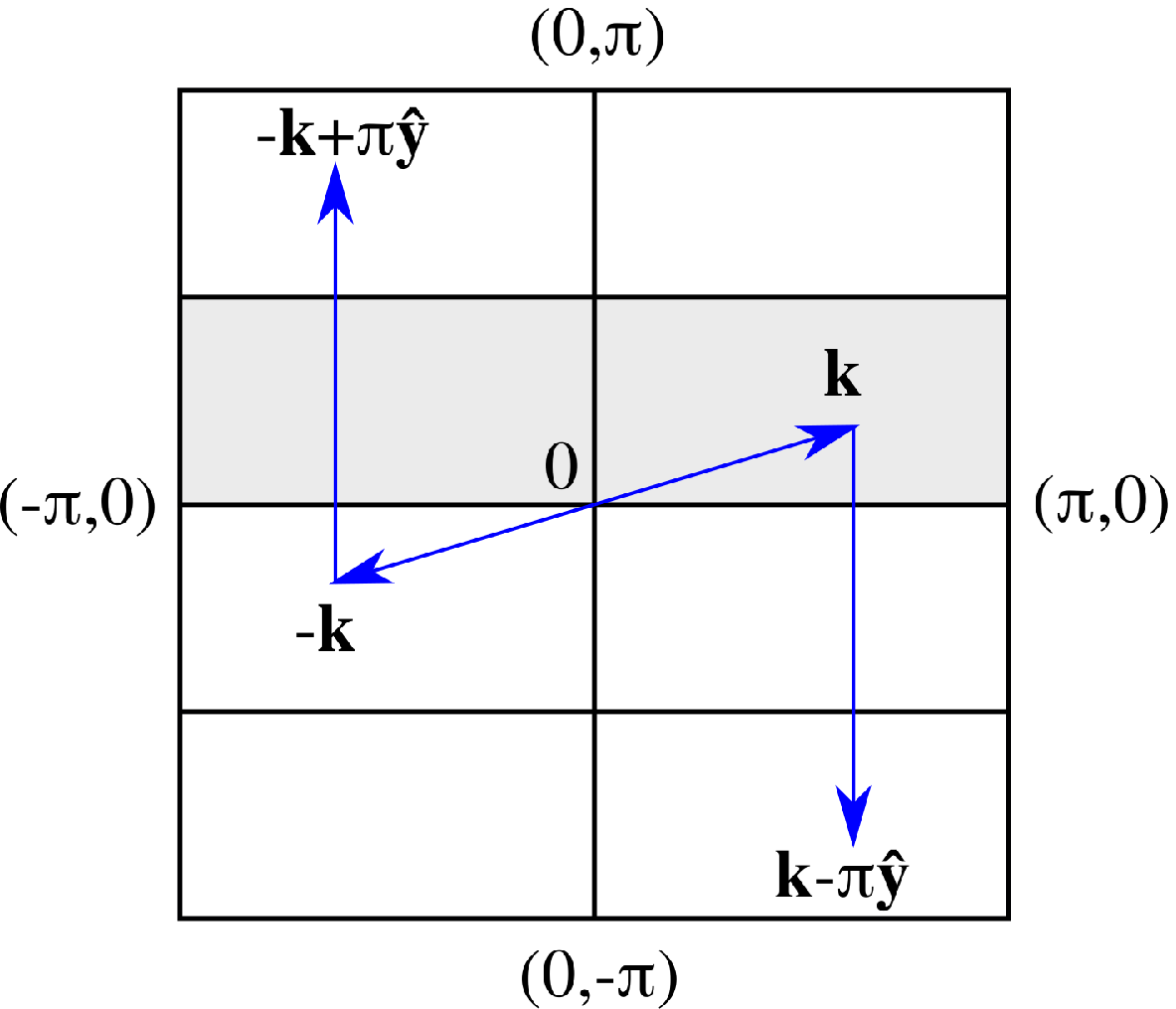}
}
\qquad
\subfloat[Zig-zag model]{
\includegraphics[angle=0,scale=0.5,clip]{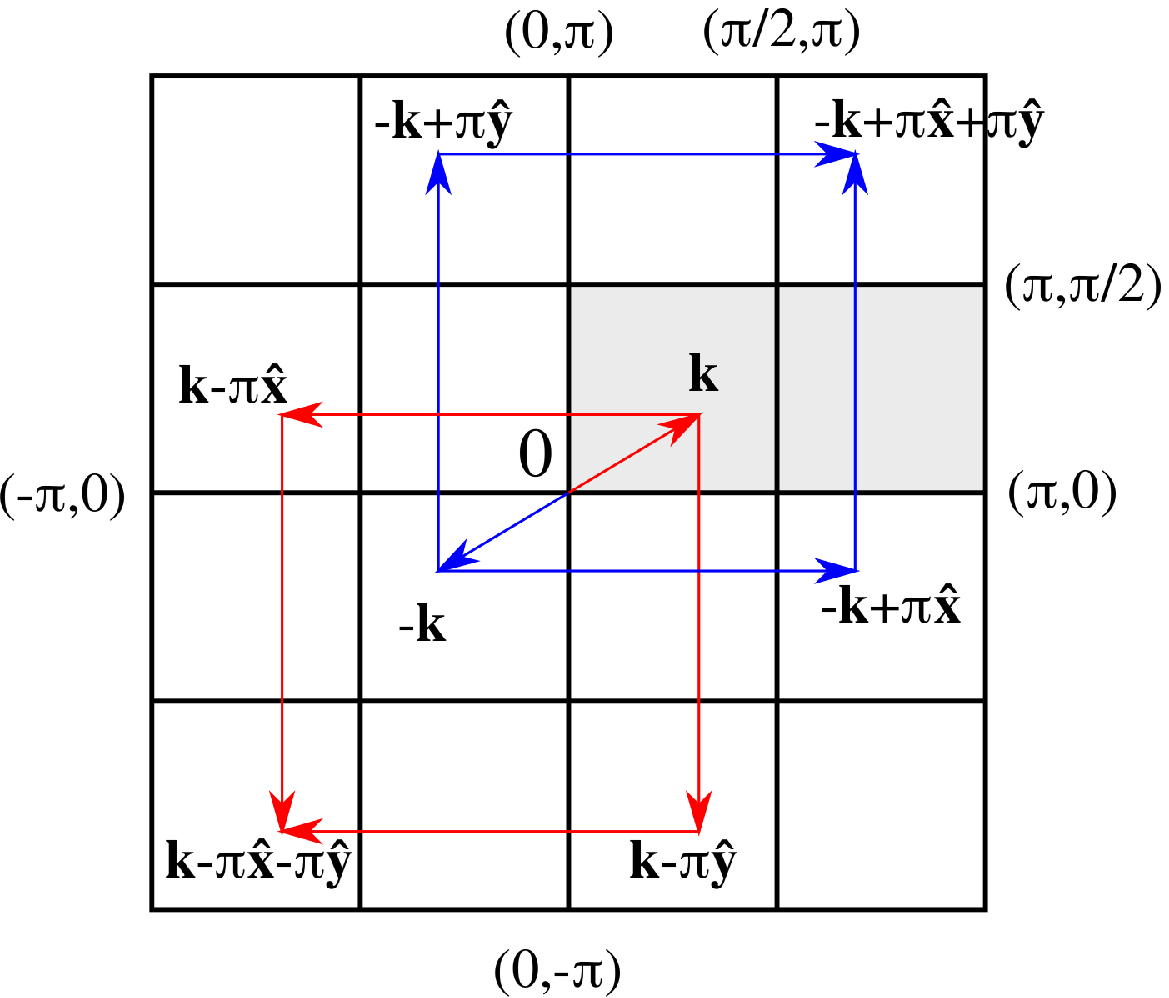}
}
\caption{\label{fig2}
Brillouin zone restricted to the domain $(k_x,k_y)$ (grey area) such that $-\pi\le k_x\le \pi$ and $0\le k_y\le \pi/2$ for the PUD model, and $0\le k_x\le \pi$ and $0\le k_y\le \pi/2$ for the ZZD model. When the vector $\Bk$ sweeps over the grey area, the other vectors fill the complementary
Brillouin zone.}
\end{figure}

\begin{figure}[!ht]
\centering
\includegraphics[angle=0,scale=0.4,clip]{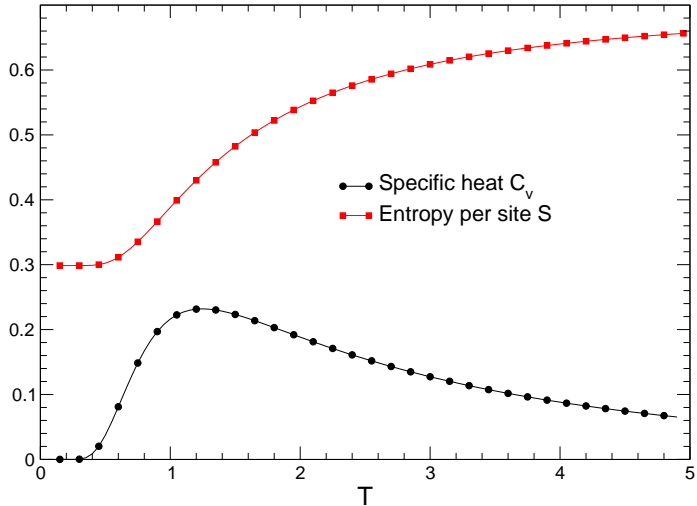}
\caption{\label{fig3}
Entropy per site and specific heat as function of temperature for the PUD model ($L=512$). The
thermodynamical functions are also valid for the ZZD model since the expression of the free energy is equivalent, see text.}
\end{figure}

%
We consider for simplification the PUD case where $\epsilon_{m,n}=(-1)^n$, see figure \ref{fig1}(a), and use the Fourier decomposition $\ar=L^{-1}\sum_{\Bk}\ak\e^{i\Bk.\Br}$, $\abr=L^{-1}\sum_{\Bk}\abk\e^{-i\Bk.\Br}$ for $\Br=(m,n)$ and $\Bk=(k_x,k_y)=(2\pi p/L,2\pi q/L)$. Using the identity $\sum_{n=0}^{L-1}e^{2i\pi qn/L}(-1)^n=L\delta_{q,\pm L/2}$, we obtain after some algebra
\bb\nn\fl
\SF=\sum_{k_x,k_y} a_{k_x,k_y}\bar a_{k_x,k_y}+u \Big (
a_{k_x,k_y}a_{-k_x,-k_y\pm \pi}e^{ik_x}+\bar a_{k_x,k_y}a_{k_x,k_y\pm \pi}e^{-ik_x}\Big )
\\ \fl
-u\Big (
a_{k_x,k_y}\bar a_{k_x,k_y}e^{ik_y}+\bar a_{k_x,k_y}\bar a_{-k_x,-k_y}e^{-ik_y}\Big )
+u^2 a_{k_x,k_y}\bar a_{k_x,k_y\pm \pi}e^{i(k_y-k_x)}.
\ee
We can decompose the momentum zone by considering the reduced domain $-\pi\le k_x\le \pi$ and $0\le k_y\le \pi/2$, see figure \ref{fig2}(a), of independent momenta. This implies that only the following set of 4 pairs of independent Grassmann variables is necessary to decompose the partition function into independent blocks
\bb\nn
a_{\Bk}=a_{k_x,k_y},\;a_{-\Bk}=a_{-k_x,-k_y},\;b_{\Bk}=a_{k_x,k_y-\pi},\;b_{-\Bk}=a_{-k_x,-k_y+\pi},
\\
\bar a_{\Bk}=\bar a_{k_x,k_y},\;\bar a_{-\Bk}=\bar a_{-k_x,-k_y},\;\bar b_{\Bk}=\bar a_{k_x,k_y-\pi},
\;\bar b_{-\Bk}=\bar a_{-k_x,-k_y+\pi}.
\ee
%
%
In the reduced grey zone of figure \ref{fig2}(a), and represented by a prime symbol on the sum over momenta, we obtain the following decomposition of the action into independent blocks of 8 variables, after regrouping the different terms and further simplifications
%
%
\bb
\nn
\SF=&{\sum_{k_x,k_y}}^{'} a_{\Bk}\bar a_{\Bk}(1-ue^{ik_y})+a_{-\Bk}\bar a_{-\Bk}(1-ue^{-ik_y})
+b_{\Bk}\bar b_{\Bk}(1+ue^{ik_y})+b_{-\Bk}\bar b_{-\Bk}(1+ue^{-ik_y})
\\ \nn
&+2iu\sin k_y \bar a_{\Bk}\bar a_{-\Bk}-2iu\sin k_y \bar b_{\Bk}\bar b_{-\Bk}
+a_{\Bk}\Big [ 2iu\sin k_xb_{-\Bk}-\bar b_{\Bk}ue^{-ik_x}(1-ue^{ik_y})\Big ]
\\ \nn
&+a_{-\Bk}\Big [-2iu\sin k_x b_{\Bk}-\bar b_{-\Bk}ue^{ik_x}(1-ue^{-ik_y})\Big ]
\\ \label{S_TF}
&+\bar a_{\Bk}b_{\Bk}ue^{-ik_x}(1+ue^{ik_y})+\bar a_{-\Bk}b_{-\Bk}ue^{ik_x}(1+ue^{-ik_y}).
\ee
The total partition function is therefore equal to the following set of integrals, where the 
$L\times L/2$ vectors $\Bk$ are located inside the reduced zone only
\bb
\ZF=2^N\cosh(K)^{2N}{\prod_{\Bk}}^{'}
\int d\bar a_{\Bk}d a_{\Bk}d\bar a_{-\Bk}d a_{-\Bk}d\bar b_{\Bk}d b_{\Bk}
d\bar b_{-\Bk}d b_{-\Bk}e^{\SF}.
\ee
We obtain after integration and simplifications
\bb\fl
\ZF=2^N\cosh(K)^{2N}(1-u^2)^{N/2}{\prod_{k_x,k_y}}^{'}\Big [
(1+u^2)^2-2u^2(\cos(2k_x)+\cos(2k_y))\Big ].
\ee
%
%
This result can be further simplified by noting that $1-u^2=\cosh^{-2}(K)$, and we finally obtain
a simple product
\bb
\ZF=2^N{\prod_{k_x,k_y}}^{'}\left [
1+\sinh^2(2K)\left (\sin^2k_x+\sin^2k_y
\right )\right ].
\ee
In the thermodynamical limit, the free energy per site is then equal to
\bb\label{F_PUD}
-\beta F/N=\ln 2+\frac{1}{\pi^2}\int_0^{\pi/2}dk_x\int_0^{\pi/2}dk_y
\ln\left [1+\sinh^2(2K)\Big (\sin^2k_x+\sin^2k_y\Big )\right ],
\ee
which corresponds to the result (6) of reference \cite{Andre:1979}. The entropy is given by $S=-\partial F/\partial T$ and the specific heat by $C_v=T\partial S/\partial T$.
In the limits $T\rightarrow 0$ and $L\rightarrow \infty$, a series expansion gives the entropy per site
\bb\nn
\lim_{T\rightarrow 0}\lim_{N\rightarrow \infty}\frac{S}{N}
\simeq \frac{1}{\pi^2}\int_{0}^{\pi/2}dk_x\int_{0}^{\pi/2}dk_y\ln\left [4\left (\sin^2 k_x+\sin^2 k_y\right )\right ]=\frac{G}{\pi}\simeq 0.291\;561,
\ee
where $G$ is the Catalan constant. The entropy is non zero as the ground state is macroscopically
degenerate. In figure \ref{fig3} we have plotted the specific heat and entropy per site.
The specific heat presents no phase transition but a Schottky peak around $T\simeq J$ as expected
for a gaped system.
%
%
We can notice that for the ZZD model, equivalent results are found. We obtain the same expression
for the free energy as equation \eref{F_PUD} except that the terms $\sin^2k_x$ are replaced
by $\cos^2k_x$. This does not change the overall expression of the free energy since we 
can always perform the change of variables $k_x\rightarrow \pi/2-k_x$ within the domain of allowed momenta. In addition, since the function $\epsilon_{m,n}$ of the ZZD model depends on both $m$ and $n$, the subsequent calculations imply that the momenta are restricted in the grey area of the Brillouin
zone represented in figure \ref{fig2}(b), with the introduction of 16 variables for each block. However each block can be divided into two independent blocks of 8 variables and therefore
the calculations are similar to the PUD case and can be simplified accordingly.

\section{Coupling with fermions \label{sect_coupling}}
%
We would like to construct the total partition function $Z=\Tr\e^{-\beta H}\simeq\int \e^{\Seff}$, where the effective action $\Seff$ includes the Ising part and the electron kinetic term in \eref{Ham}. This can be reformulated as a Grassmannian functional integral by substituting the Fermi operators with dimensionless Grassmann variables $\ccr{r}\rightarrow \cg{\Br}(\tau)$ and $\cdr{r}\rightarrow \cbg{r}(\tau)$, see for example the technical details in references \cite{book:altland:10,lichtenstein:17}, where $\tau$ is the imaginary time variable taking values between 0 and $\beta$
\bb\nn\fl
Z=&2^N\cosh(K)^{2N}\int \prod_{\Br}\drm \abr\drm \ar\drm b_{\Br}
\int \prod_{\Br,\tau} \drm \cbg{r}(\tau)\drm \cg{r}(\tau)
\\ \nn
&\exp\left (
\SF+\sum_{\Br}b_{\Br}\left [\frac{1}{2}-\beta^{-1}\int_0^{\beta}\drm\tau\cbg{r}(\tau)\cg{r}(\tau)\right ]
+\sum_{\Br}b_{\Br}
\left [
\frac{1}{2}+\beta^{-1}\int_0^{\beta}\drm\tau\cbg{r}(\tau)\cg{r}(\tau)\right ]\ar\abr 
\right .
\\ \label{Stot}
&+\left .\sum_{\Br}\int_0^{\beta}\drm\tau \left [
-\cbg{r}(\tau)\partial_{\tau}\cg{r}(\tau)+t\sum_{\Bd}\cbg{r+\Bd}(\tau)\cg{r}(\tau)
\right ] \right ).
\ee
The additional nilpotent variables $b_{\Br}$ in the integral are introduced in order to impose the hole constraints inside equation \eref{Z_int}. Instead of a nilpotent variable, a pair of Grassmann variables could also have been introduced as well. The following prescription was considered whenever the product of two Fermi operators on the same site are encountered: $\cdr{r}\ccr{r}=(\cdr{r}\ccr{r}-\ccr{r}\cdr{r}+1)/2\rightarrow \cbg{\Br}(\tau)\cg{\Br}(\tau)+1/2$. Moreover,
we assume that only the quadratic terms $a_{\Br}\bar a_{\Br}$ in equation \eref{Int} contribute to the interaction between the spins and fermions as discussed previously. This will simplify the calculations
and the integrals are reduced to Pfaffians or determinants.
In the following section we study the extrema of this action using Lagrange multipliers in the mean-field approximation. From the action $\Seff$ defined in equation \eref{Stot}, we impose  antisymmetric boundary conditions $\cg{r}(0)=-\cg{r}(\beta)$ and $\cbg{r}(0)=-\cbg{r}(\beta)$. These operators can be expressed as a Fourier series using Matsubara frequencies $\omega_n=(2n+1)\pi/\beta$, $n=0,\pm 1,\pm 2,\cdots$, such that
\bb
\cg{r}(\tau)=\beta^{-1/2}\sum_{\omega_n}\e^{-i\omega_n\tau}
\cg{r}(\omega_n),\;
\cbg{r}(\tau)=\beta^{-1/2}\sum_{\omega_n}\e^{i\omega_n\tau}
\cbg{r}(\omega_n).
\ee
To simplify the problem, we observe that the functional \eref{Stot} depends on two types of fluctuating operators, coming from the representation of the spins and fermions, with a coupling between them that depends on the local hole constraints. The spin fluctuations are described by the quadratic part with the pairs $(\ar,\abr)$, and the dynamics of the fermions also by a quadratic part with the time dependent variables $(\cg{r}(\tau),\cbg{r}(\tau))$. Integrating over $(\ar,\abr)$
and the coupling operators $b_{\Br}$ introduces nonlocal interactions between the fermions and is not 
treatable exactly because of the strong vacancy constraints since the density operators in factor of $b_{\Br}$ have only zero or one as possible values. This strong condition can be lifted if we consider local mean-field density for the fermion sector. In that case, the singularities coming from the $b_{\Br}$ couplings are smoothed out and self-consistent local mean-field equations can be expressed in function of the local densities and self-energies of the fermions.
Under these assumptions, we replace the operators $\cbg{r}(\omega_n)\cg{r}(\omega_n)$ by a Green function $G_{\Br}(\omega_n)$ with conjugate variable $\Sigma_{\Br}(\omega_n)$, or self-energy, which serves as a Lagrange multiplier. After integration over the $b_{\Br}$ variables, the total effective action, including the spin and fermion parts with their interaction is finally given by a quadratic Grassmannian depending on $(a_{\Br},\bar a_{\Br})$ and $(\cg{r}(\omega_n),\cbg{r}(\omega_n))$
\bb\nn
\Seff&=N\ln 2+2N\ln\cosh(K)+\SF+\sum_{\Br}\ln\left (\frac{1}{2}-\beta^{-1}\sum_{\omega_n}G_{\Br}(\omega_n)\right )
\\ \nn
&+\sum_{\Br}\left .\frac{1+2\beta^{-1}\sum_{\omega_n}G_{\Br}(\omega_n)}
{1-2\beta^{-1}\sum_{\omega_n}G_{\Br}(\omega_n)}\right .\ar\abr
+\sum_{\Br,\omega_n}\Sigma_{\Br}(\omega_n)\Big [G_{\Br}(\omega_n)-\cbg{r}(\omega_n)\cg{r}(\omega_n) \Big ]
\\ \label{Seff}
&+\sum_{\Br}\sum_{\omega_n} \left [
(i\omega_n+\mu)\cbg{r}(\omega_n)\cg{r}(\omega_n)+t\sum_{\Bd}\cbg{r+\Bd}(\omega_n)\cg{r}(\omega_n)
\right ].
\ee
In this expression, we have introduced a chemical potential $\mu$. The second term in the second
line corresponds to the local constraints imposed by the Lagrange multipliers. The effective thermodynamical potential $-\beta\Omega=\ln(\int \e^{\Seff})$ after integration
over the Grassmann variables is given by
\bb\nn
-\beta\Omega=N\ln 2+2N\ln \cosh (K)+\sum_{\Br}\ln\left (\frac{1}{2}-\beta^{-1}\sum_{\omega_n}G_{\Br}(\omega_n)\right )
+\sum_{\Br,\omega_n}\Sigma_{\Br}(\omega_n)G_{\Br}(\omega_n)
\\ \label{Omega_gen}
+\ff\ln \det \Big (\AF+V \Big )_{\Br,\Br'}
+\sum_{\omega_n}\ln\det 
\Big ([i\omega_n+\mu-\Sigma_{\Br}(\omega_n)]\delta_{\Br,\Br'}+t\sum_{\Bd}\delta_{\Br',\Br+\Bd}
\Big )_{\Br,\Br'},
\ee
where $\AF$ is the connectivity matrix corresponding to the spin sector in the real space, see equation \eref{SF}: $\SF=\ff\sum_{\Br,\Br'}\Psi_{\Br}^{\trans} (\AF)_{\Br,\Br'}\Psi_{\Br'}$, in the basis of vectors $\Psi_{\Br}=(\ar,\abr)^{\trans}$, and $V$ the operator corresponding to the hole coupling at the quadratic approximation
\bb\nn
(V)_{\Br,\Br'}=\frac{1+2\beta^{-1}\sum_{\omega_n}G_{\Br}(\omega_n)}
{1-2\beta^{-1}\sum_{\omega_n}G_{\Br}(\omega_n)}i\sigma_2\delta_{\Br,\Br'}=
v_{\Br}i\sigma_2\delta_{\Br,\Br'},
\\ 
v_{\Br}=\frac{n_{\Br}}{1-n_{\Br}},\;
n_{\Br}=\ff+\beta^{-1}\sum_{\omega_n}G_{\Br}(\omega_n).
\ee
%
\subsection{Homogeneous solution \label{sect_homog}}
We consider the mean field approach of the effective action \eref{Seff} with an homogeneous solution for the Green function and self energy: $G_{\Br}(\omega_n)=G(\omega_n)$, $\Sigma_{\Br}(\omega_n)=\Sigma$. This implies that $v_{\Br}=v=n_e/(1-n_e)$ where $n_e=v/(1+v)$ is the fermion average density. By adding the action $\SF$ \eref{S_TF} and the quadratic interaction part of equation \eref{S_int}, after discarding the quartic terms, we find that all the contributions depending on $(a_{\Br},\bar a_{\Br})$ are included in the following action
\bb\label{Sint_F}
\Sint=
\SF+v\sum_{m,n}a_{m,n}\bar a_{m,n}-N\ln (1+v).
\ee
%
%
We can express the interaction part using Fourier modes in the reduced Brillouin zone of figure 
\ref{fig2}(a)
\bb
v\sum_{m,n}a_{m,n}\bar a_{m,n}=v{\sum_{k_x,k_y}}^{'} \Big ( a_{\Bk}\bar a_{\Bk}
+a_{-\Bk}\bar a_{-\Bk}+b_{\Bk}\bar b_{\Bk}+b_{-\Bk}\bar b_{-\Bk} \Big ),
\ee
and we obtain after integration $\ZF(v)=2^N\cosh(K)^{2N}\int e^{\Sint}$, where
\bb\label{Zv}
\ZF(v)=&\frac{2^N\cosh(K)^{2N}}{(1+v)^N}{\prod_{k_x,k_y}}^{'}\Big [
(1-u^4)^2+(1+v)^4-4vu^4-1 
\\ \nn
&+2u^4v^2\cos 2(k_y-k_x)
-2u^2(1+v-u^2)^2 \Big (\cos 2k_x+\cos 2k_y\Big )\Big ].
\ee
The Grand Potential given by \eref{Omega_gen} can be written in the mean-field approximation as
\bb\fl \nn
-\frac{\beta}{N}\Omega&=\ln 2+2\ln \cosh K+
\ln(1-n_e)+\left (n_e-\ff\right )\beta\Sigma
\\ \fl\label{Omega_saddle}
&+\frac{1}{N}\sum_{\Bk}\ln \cosh\left [\frac{\beta}{2}\left (\mu-\Sigma-\epsilon_{\Bk}\right )\right ]
+\frac{1}{N}{\sum_{\Bk}}^{'}\ln\left [(1-u^4)^2+(1+v)^4-4vu^4-1 \right .
\\ \nn
&\left .+2u^4v^2\cos 2(k_y-k_x)
-2u^2(1+v-u^2)^2\Big (\cos 2k_x+\cos 2k_y\Big )\right ].
\ee
The self-consistent equations are obtained by extremization of \eref{Omega_gen} with respect
to the Green functions $G(\omega_n)$ and self energy $\Sigma$. This yields the two equations
\bb\label{homog_sol}
\beta\Sigma(\mu)=\frac{1}{1-n_e}-\frac{\FF(v)}{(1-n_e)^2},
\\
n_e=\ff+\beta^{-1}\sum_{\omega_n}G(\omega_n),\;G(\omega_n)=\frac{1}{N}\sum_{\Bk}
\frac{1}{i\omega_n+\mu-\Sigma-\epsilon_{\Bk}}.
\ee
In the last expression, we have redefined the kinetic
energy of the fermions by $\epsilon_{\Bk}=-t(\cos k_x+\cos k_y-2)$ by shifting $\mu$. Function $\FF(v)$
is given by 
\bb\fl\fl
\FF(v)=\frac{4}{N}{\sum_{\Bk}}^{'}
\frac{(1+v)^3-u^4+u^4v\cos 2(k_y-k_x)-u^2(1+v-u^2)\Big (\cos 2k_x+\cos 2k_y\Big )}{(1-u^4)^2+(1+v)^4-4vu^4-1+2u^4v^2\cos 2(k_y-k_x)
-2u^2(1+v-u^2)^2\Big (\cos 2k_x+\cos 2k_y\Big )}.
\ee
%
\begin{figure}[!ht]
\centering
\includegraphics[angle=0,scale=0.4,clip]{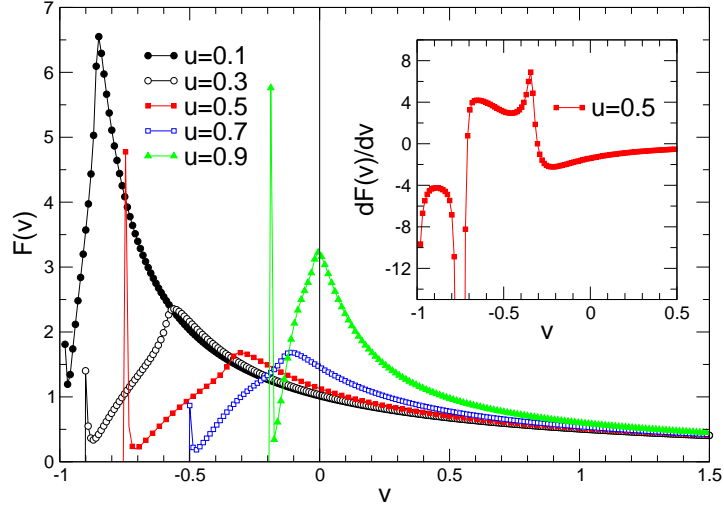}
\caption{\label{fig4}
Function $\FF(v)$ for different temperatures ($L=64$). The physical domain corresponds to $v>0$.
All the singularities are located in the domain $v\le 0$.}
\end{figure}
%
Function $\FF(v)$ is plotted in figure \ref{fig4} for different temperatures. Physically, $\FF(v)$
is defined for $v\ge 0$, whereas for $v<0$ (negative densities) the function presents different singularities which are not real. However we have represented these singularities for an illustration
purpose only, as the mathematical domain of $\FF(v)$ is not restricted to positive values. $\FF(v)$ can be viewed as a local chemical potential due to the spin environment and proportional to the fermion self-energy. This quantity is also related to the energy to insert a hole inside the spin bath. In the limit of low density of fermions $v\simeq 0$, we have indeed $\Sigma(\mu)\simeq k_BT(1-\FF(0))$ which is not zero and is temperature dependent. When $T=0$, a singularity appears at $v=0$ and has the shape of a cusp, see the green curve for $T$ close to zero ($u=0.9$). This corresponds to the phase transition of the non-dilute frustrated model at $T=0$.
We estimate the effective mass from the derivative of $\Sigma$ with respect to the chemical potential using the definition $m^*=1-\partial\Sigma/\partial\mu$ \cite{book:Mahan,Asgari:2005}. $\Sigma'(\mu)$ can be computed and expressed with $\FF(v)$ and its derivative $\FF'(v)=\partial_v\FF(v)$
\bb\label{Lambda}
\beta \frac{\partial \Sigma}{\partial \mu}=\frac{n_e'}{(1-n_e)^4}\left [
(1-n_e)^2-2(1-n_e)\FF(v)-\FF'(v) \right ]
=n_e'\Lambda,
\ee
with the density of states $n'_e=\partial_{\mu}n_e$ defined by
\bb
n_e'=\left (1-\frac{\partial \Sigma}{\partial \mu}\right )
\frac{\beta}{4N}\sum_{\Bk}
\left [1-\tanh^2\fdfrac{\beta}{2}(\mu-\Sigma-\epsilon_{\Bk})\right ]
=\left (1-\frac{\partial \Sigma}{\partial \mu}\right )\frac{\beta \tilde n_e'}{4},
\ee
where $0<\tilde n_e'<1$ for any temperature. It is known that the density of states is proportional
to the effective mass. We can finally express the effective mass as $m^*=(1+\tilde n_e'\Lambda/4)^{-1}$. In the high temperature regime, we can perform a series expansion for $u\ll 1$ of function $\FF(v)$
\bb
\FF(v)\simeq \frac{1}{1+v}+\frac{3u^4}{(1+v)^4},
\ee
and obtain an estimate of the self-energy, $\Sigma\simeq -3(1-n_e)^2J^3/(k_BT)^3<0$, from which
\bb
m^*\simeq 1-\frac{3}{2}u^4(1-n_e)<1.
\ee
For the non-frustrated Ising model, $\epsilon_{\Br}=1$, it is worth noting that we obtained instead the opposite sign for this approximation in the same limit $m^*\simeq 1+\frac{3}{2}u^4(1-n_e)>1$
\cite{Fortin:2021}.
\subsection{Solution with stripes \label{sect_stripes}}
The previous solution does not incorporate the symmetry of the spin couplings. We 
expect that the charge density will follow a modular structure similar to the
local couplings $J\epsilon_{\Br}$.
We consider therefore the set of trial mean-field solutions with alternating values, $n_{m,n}=n_1$ if $n$ is even and $n_{m,n}=n_2$ otherwise when $n$ is odd
\bb
n_{m,n}=\ff n_1(1+(-1)^n)+\ff n_2(1-(-1)^n).
\ee
We also define the quantities $v_1=n_1/(1-n_1)$ and $v_2=n_2/(1-n_2)$. The quadratic interaction part of the action \eref{Sint_F} can be decomposed as
\bb\nn
\sum_{m,n}v_{m,n}a_{m,n}\bar a_{m,n}&=
\frac{v_1+v_2}{2}{\sum_{k_x,k_y}}^{'} \Big ( a_{\Bk}\bar a_{\Bk}
+a_{-\Bk}\bar a_{-\Bk}+b_{\Bk}\bar b_{\Bk}+b_{-\Bk}\bar b_{-\Bk} \Big )
\\
&+\frac{v_1-v_2}{2}{\sum_{k_x,k_y}}^{'} \Big ( a_{\Bk}\bar b_{\Bk}
+a_{-\Bk}\bar b_{-\Bk}+b_{\Bk}\bar a_{\Bk}+b_{-\Bk}\bar a_{-\Bk} \Big ).
\ee
After including these terms to $\SF$, we obtain the following partition function for the frustrated Ising part that includes the fermion coupling
\bb\fl\fl\nn
\ZF(v_1,v_2)=&\frac{2^N\cosh(K)^{2N}}{(1+v_1)^{N/2}(1+v_2)^{N/2}}
{\prod_{\Bk}}^{'}
\Big [
(1-u^4)^2+(1+v_1)^2(1+v_2)^2-2(v_1+v_2)u^4-1 
\\ \nn\fl
&+u^2(v_1-v_2)^2+2u^4v_1v_2\cos 2\left (k_y-k_x\right )
-2u^2(1+v_1-u^2)(1+v_2-u^2)\Big (\cos 2k_x
+\cos 2k_y\Big )
\\ \nn\fl
&
-2u(v_1-v_2)(u^2-(1+v_1)(1+v_2))\cos\left (k_x\right )
+2u^3(u^2-1)(v_1-v_2)\cos\left (k_x-2k_y\right )
\Big ]
\\
=&\frac{2^N\cosh(K)^{2N}}{(1+v_1)^{N/2}(1+v_2)^{N/2}}\QF(v_1,v_2).
\ee
When $v_1=v_2=v$, the last two terms in the product vanish and we recover equation \eref{Zv},
$\ZF(v,v)=\ZF(v)$. The space asymmetry is revealed by the terms proportional to $(v_1-v_2)$.
The local self-energies and Green functions are expressed similarly
\bb\nn
\Sigma_{\Br}(\omega_n)=\ff \Sigma_1(\omega_n)(1+(-1)^n)+\ff \Sigma_2(\omega_n)(1-(-1)^n)
=S_1(\omega_n)+(-1)^nS_2(\omega_n),
\\
G_{\Br}(\omega_n)=\ff G_1(\omega_n)(1+(-1)^n)+\ff G_2(\omega_n)(1-(-1)^n),
\ee
and the determinant $\Delta_{\omega_n}(\Sigma_1,\Sigma_2)$ of the fermionic kinetic part defined by
\bb
\Delta_{\omega_n}(\Sigma_1,\Sigma_2)=\det 
\Big ([i\omega_n+\mu-\Sigma_{\Br}(\omega_n)]\delta_{\Br,\Br'}+t\sum_{\Bd}\delta_{\Br',\Br+\Bd}
\Big )_{\Br,\Br'},
\ee
can be evaluated by considering the eigenvalues $\lambda$ of the corresponding matrix in the Fourier space. If we note $\varphi_{\Bk}$ the eigenvectors of $\lambda$, we have the set of secular equations
\bb\nn
\Big [i\omega_n+\mu-S_1-\epsilon_{\Bk}-\lambda \Big ]\varphi_{\Bk}-S_2\bar\varphi_{\Bk}=0,
\\
S_2\varphi_{\Bk}-\Big [i\omega_n+\mu-S_1-\bar\epsilon_{\Bk}-\lambda \Big ]\bar\varphi_{\Bk}=0,
\ee
where $\bar\varphi_{\Bk}=\varphi_{\Bk=(k_x,k_y\pm \pi)}$, and similarly for $\bar\epsilon_{\Bk}=\epsilon_{\Bk=(k_x,k_y\pm \pi)}$. This leads to the solutions
\bb
\lambda=\lambda_{\Bk}^{\pm}(\omega_n)=i\omega_n+\mu-S_1-\ff(\epsilon_{\Bk}+\bar\epsilon_{\Bk})
\pm\ff\sqrt{\left (\epsilon_{\Bk}-\bar\epsilon_{\Bk}\right )^2+4S_2^2},
\ee
and the determinant $\Delta_{\omega_n}$ is directly equal to
\bb
\Delta_{\omega_n}(\Sigma_1,\Sigma_2)=
\left ({\prod_{\Bk}}^{'}\lambda_{\Bk}^{+}(\omega_n)\lambda_{\Bk}^{-}(\omega_n)\right )^2,
\ee
where the prime symbol corresponds as before to the sum reduced to the Fourier modes of the 
Brillouin zone of figure \ref{fig2}(a), and the square takes into account the remaining modes
that correspond to negative values of $k_y$.
The Grand potential can then be expressed from equation \eref{Omega_gen} by replacing
the different parameters with the trial solution. Thus
\bb\nn
-\frac{\beta\Omega}{N}&=\ln 2+2\ln\cosh(K)+\ff\ln(1-n_1)+\ff\ln(1-n_2)
\\ \nn
&+\ff\sum_{\omega_n}\Sigma_1(\omega_n)G_1(\omega_n)+
\ff\sum_{\omega_n}\Sigma_2(\omega_n)G_2(\omega_n)
\\ \label{Omega_frust}
&+\frac{1}{N}\ln \QF(v_1,v_2)
+\sum_{\omega_n}\frac{2}{N}{\sum_{\Bk}}^{'}
\ln\left ( \lambda_{\Bk}^{+}(\omega_n)\lambda_{\Bk}^{-}(\omega_n)\right ).
\ee
From this expression, we obtain the mean-field solutions by considering the different derivatives
with respect to the quantities $\{\Sigma_1(\omega_n),G_1(\omega_n),\Sigma_2(\omega_n),G_2(\omega_n)\}$, which leads to the following expressions for the pair of Green functions
%
%
\bb\nn
G_1(\omega_n)=\frac{2}{N}{\sum_{\Bk}}^{'}\sum_{\sigma=\pm 1}\frac{1}{\lambda_{\Bk}^{\sigma}(\omega_n)}\left (1-\frac{2\sigma S_2}{\sqrt{\left (\epsilon_{\Bk}-\bar\epsilon_{\Bk}\right )^2+4S_2^2}}\right ),
\\
G_2(\omega_n)=\frac{2}{N}{\sum_{\Bk}}^{'}\sum_{\sigma=\pm 1}\frac{1}{\lambda_{\Bk}^{\sigma}(\omega_n)}\left (1+\frac{2\sigma S_2}{\sqrt{\left (\epsilon_{\Bk}-\bar\epsilon_{\Bk}\right )^2+4S_2^2}}\right ).
\ee
%
%
\begin{figure}[!htb]
\centering
\includegraphics[angle=0,scale=0.4,clip]{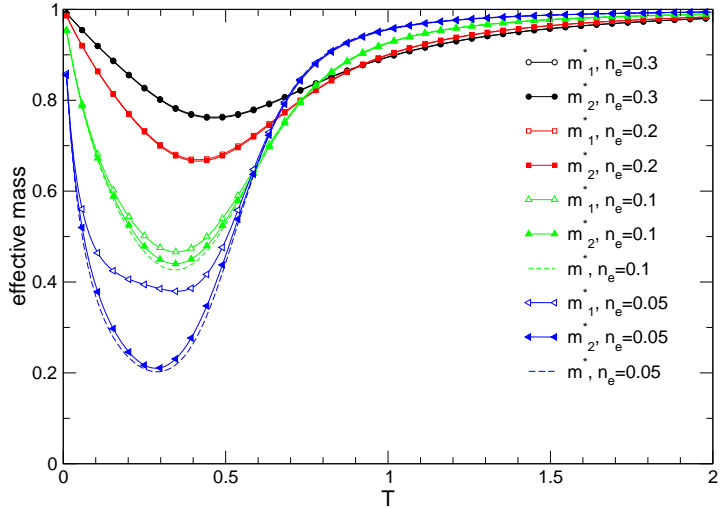}
\caption{\label{fig5}
Effective masses ${m_1}^*$ and ${m_2}^*$ as function of the temperature for different densities $n_e$ ($L=150$). The dotted lines represent the solution $m^*$ in the homogeneous case.}
\end{figure}
%
%
%
\begin{figure}[!htb]
\centering
\includegraphics[angle=0,scale=0.5,clip]{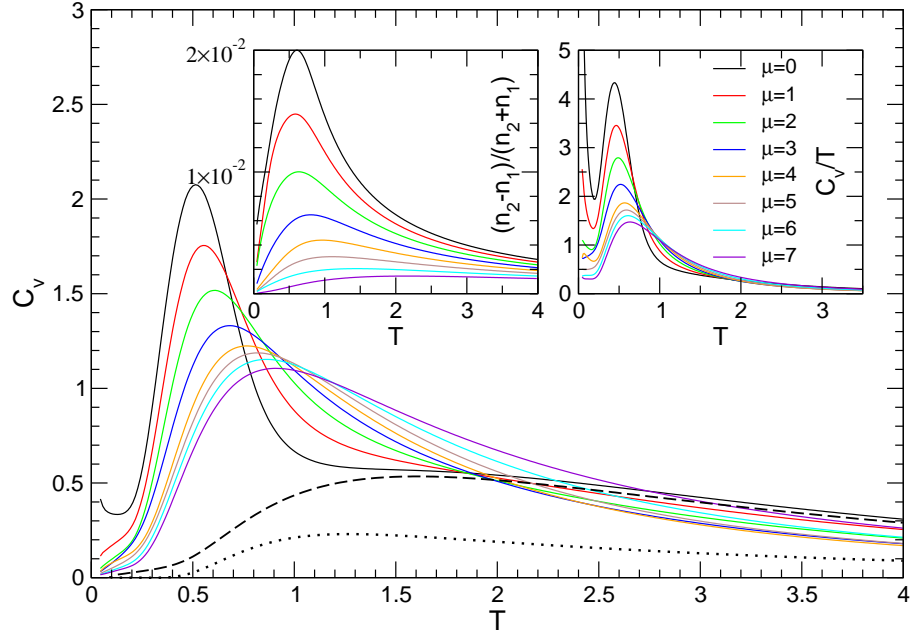}
\caption{\label{fig6a}
Specific heat as function of the temperature for different chemical potentials ($L=200$).
The dotted line is the specific heat for the spins of the non-dilute PUD system, and the dashed line
is the sum of the previous specific heat with the specific heat of free fermions at $\mu=0$, which
corresponds to low density of particles.
At high temperature it coincides with the correlated system. In insets are represented the
relative difference of the densities, and the ratio $C_v/T$. This ratio is diverging at low
temperature for low density or low chemical potential.}
\end{figure}
%
%
\begin{figure}[!htb]
\centering
\includegraphics[angle=0,scale=0.5,clip]{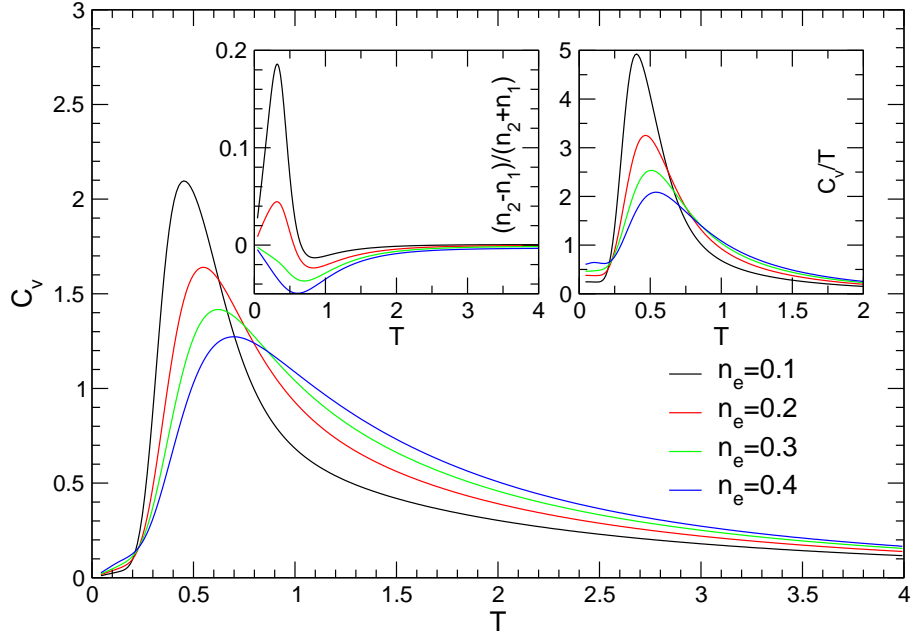}
\caption{\label{fig6b}
Specific heat in the canonical ensemble as function of the temperature for different fermion
densities ($L=200$). In insets are represented the relative difference of the densities, and the ratio $C_v/T$. This ratio tends to a constant at low temperature which decreases with density.}
\end{figure}
%
The self-energies satisfy moreover the equations
\bb\nn
\beta\Sigma_1(\omega_n)=\frac{1}{1-n_1}-\frac{2}{(1-n_1)^2}\frac{\partial}{\partial v_1}
\ln \QF(v_1,v_2),
\\
\beta\Sigma_2(\omega_n)=\frac{1}{1-n_2}-\frac{2}{(1-n_2)^2}\frac{\partial}{\partial v_2}
\ln \QF(v_1,v_2),
\ee
with the density average relation $n_1+n_2=2n_e$. They depend on the Green functions by
the identities
\bb
n_1=\ff+\beta^{-1}\sum_{\omega_n}G_1(\omega_n),\;
n_2=\ff+\beta^{-1}\sum_{\omega_n}G_2(\omega_n).
\ee
The previous homogeneous case $n_1=n_2=n_e$ is a particular solution of this set of equations.
The associated effective masses ${m_i}^{*}=1-\partial_{\mu}\Sigma_i$ are evaluated
analytically although the expressions are cumbersome. They can be derived directly from the two
following equalities involving the densities of states ${n_i}^{'}=\partial_{\mu}n_i$,
the derivatives $\FF_{i}=\partial_{v_i}\log\QF(v_1,v_2)$, and $\FF_{i,j}=\partial_{v_i}\partial_{v_j}\log\QF(v_1,v_2)$
\bb\nn
\beta\frac{\partial \Sigma_1}{\partial \mu}=
\frac{{n_1}^{'}}{(1-n_1)^4}\left [
(1-n_1)^2-4(1-n_1)\FF_1-2\FF_{1,1}
\right ]-\frac{2{n_2}^{'}}{(1-n_1)^2(1-n_2)^2}\FF_{2,1},
\\
\beta\frac{\partial \Sigma_2}{\partial \mu}=
\frac{{n_2}^{'}}{(1-n_2)^4}\left [
(1-n_2)^2-4(1-n_2)\FF_2-2\FF_{2,2}
\right ]-\frac{2{n_1}^{'}}{(1-n_1)^2(1-n_2)^2}\FF_{1,2}.
\ee
The homogeneous solution \eref{homog_sol} is recovered when $\Sigma_1=\Sigma_2=\Sigma$ by noting that $\FF'(v)=\FF_{1,1}+\FF_{1,2}+\FF_{2,1}+\FF_{2,2}$. We have plotted in figure \ref{fig5} the effective masses ${m_1}^{*}$ and ${m_2}^{*}$ as function of the average density $n_e$. Their value is always less that unity, and decreases to a minimum at low temperature. The minimum is more accentuated at low density. In the same figure we have also plotted in dashed lines the homogeneous solution $m^*$, see equation \eref{Lambda}. We observe in all the temperature regime that $m_1^*>m_2^*>m^*$ and that $\Lambda>0$. 
For a given density $n_e$, we define the free energy using the relation $F=\Omega+N\mu(n_e-1/2)$.
In figure \ref{fig6a} we have plotted the specific heat $C_v=-T\partial^2\Omega/\partial T^2$
for different chemical potentials, and in figure \ref{fig6b} the specific heat in the canonical ensemble $C_v=-T\partial^2F/\partial T^2$.
The relative density $(n_2-n_1)/(n_2+n_1)$ shows that the fermions predominantly
align along the frustrated coupling lines, depending on the initial conditions.  If we consider 
the ratio $C_v/T$ at low temperature, which, in the theory of Fermi liquids, tends to a constant proportional to the effective mass \cite{Varma:2002}, it appears that this coefficient decreases with density in the canonical ensemble, see inset of figure \ref{fig6b}. This is consistent with an effective mass decreasing with the density of fermions. However, in inset of figure \ref{fig6a}, the ratio $C_v/T$ at low
densities or for low chemical potential seems to diverge in the grand canonical ensemble, which is inconsistent with a finite entropy and probably due to the mean-field approximation. This behavior can be seen experimentally in the ground state of some spin-ice systems such as $\mathrm{Dy{Ti}_2O_7}$ \cite{Pomaranski:2013}, where a non-vanishing specific heat at zero temperature was observed, suggesting a spin-glass transition due to the increase of the spin relaxation time and macroscopic degeneracy of the ground state. This problem is not solved in our mean-field approximation, and requires a more precise analysis. However at higher temperature and above the Schottky peak, the decoupling between spins and fermions is evidenced in figure \ref{fig6a} by the dashed line which corresponds to the sum of the two independent specific heats of 
the non-dilute spin system and free fermions. The magnitude of the Schottky peak seems to suggest a strong coupling effect between the spins and fermions in low temperature regime near $T\simeq
J$.

%
\section{Numerical results with a classical model \label{sect_num}}
%
In this section we present a Monte-Carlo algorithm based on the diffusion of classical particles
on the frustrated PUD lattice. The particles in the
model \eref{Ham} are considered as classical in order to simplify the numerical difficulties inherent to Quantum Monte Carlo algorithms. We expect however qualitative results that can be applied to the quantum case.
Monte-Carlo simulations on classical particle-spin models were performed 
\cite{selke:02} to mimic the physics of "telephone-number" or ladder compounds such as cuprates
(La,Sr,Ca)$_{14}$Cu$_{24}$O$_{41+\delta}$ \cite{Golden:2001}. Their idea is to construct a spin-1 Hamiltonian with, in particular, ferromagnetic couplings between spins $\sigma_{\Br}=\pm 1$ on the horizontal lines and antiferromagnetic couplings on the vertical lines. The holes or defects are represented by $\sigma_{\Br}=0$, and the two horizontal spins surrounding the hole are connected by
one antiferromagnetic coupling. The phase diagram displays vertical lines of holes as they tends to
form stripes at low temperature. Although the model does not incorporate frustration, the appearance
of stripes seems to depend strongly on the experimental coupling values and the choice of their configurations on the lattice. The corresponding classical Hamiltonian includes quadratic spin-spin couplings as well as a quartic term that represents the presence of the itinerant holes. 
%
\begin{figure}[!htb]
\centering
\subfloat[Low temperature configuration]{
\includegraphics[angle=0,scale=0.4,clip]{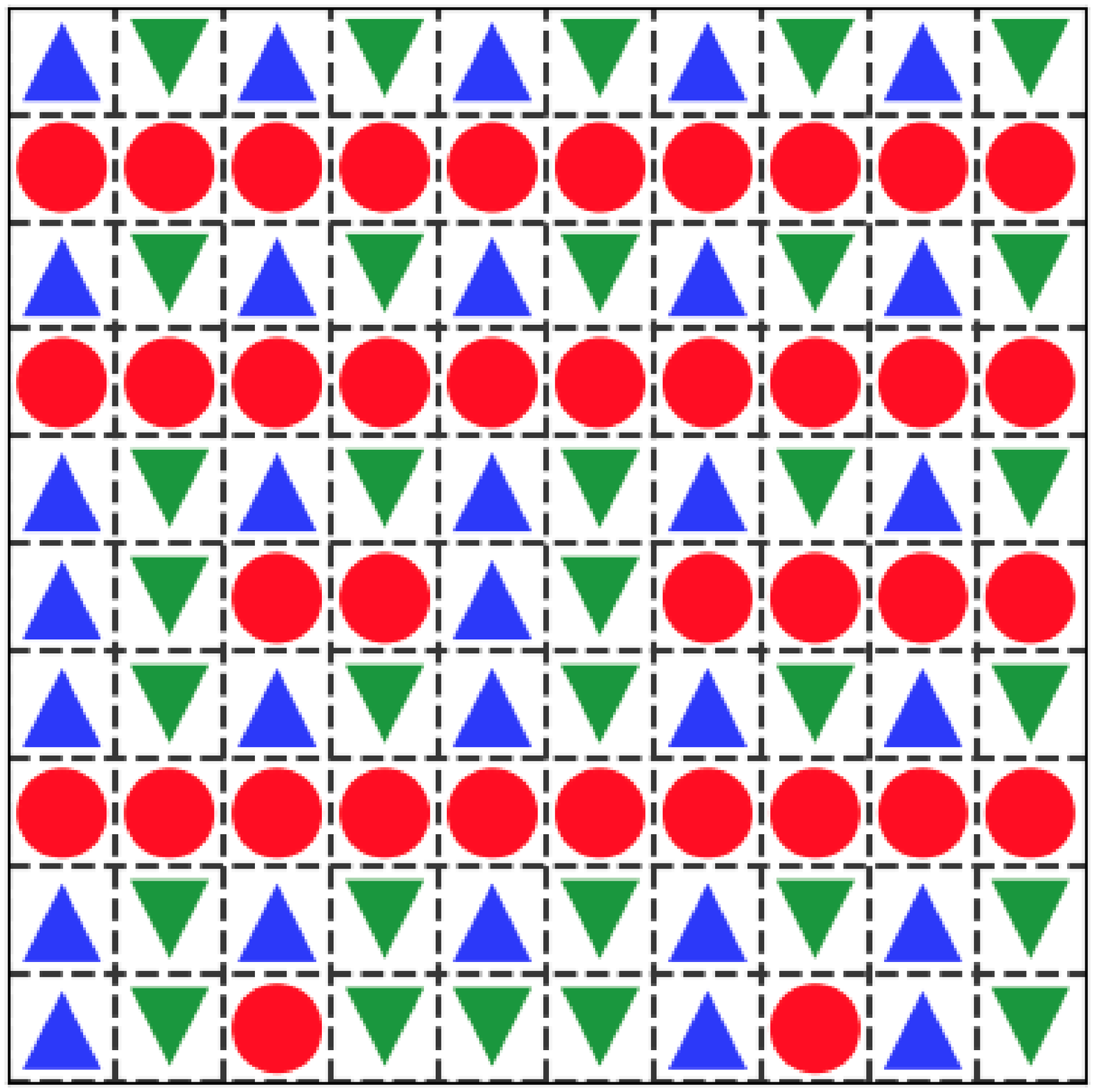}
}
\qquad
\subfloat[High temperature configuration]{
\includegraphics[angle=0,scale=0.4,clip]{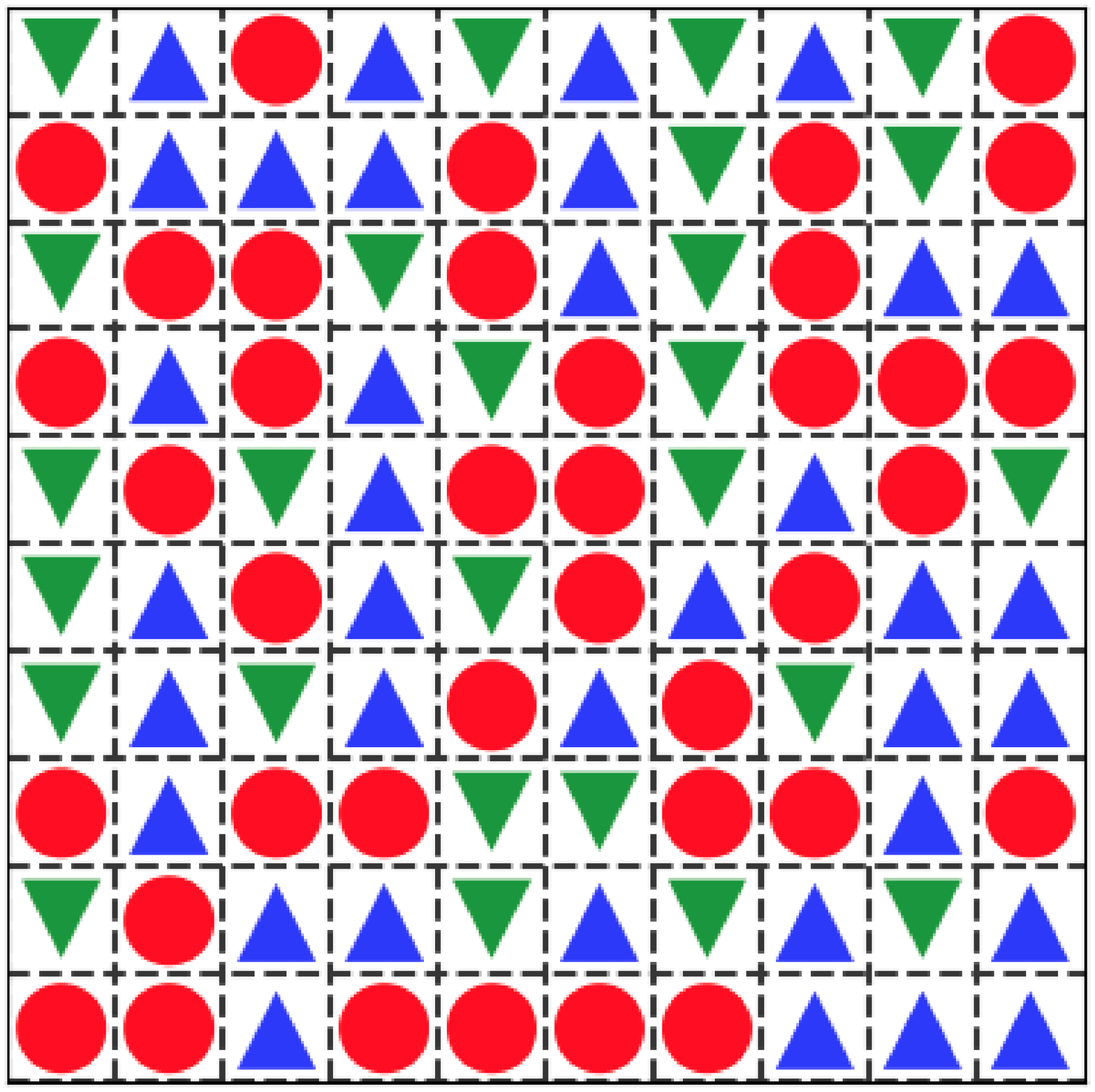}
}
\caption{\label{fig8}
Two MC final configurations: $T=0.1$ (a) and $T=0.36$ (b), for $N=100$ sites and hole density $n_e=0.4$. The spins up or down are represented by blue and green arrows respectively. The red disks
represent the vacancies.}
\end{figure}
%
\begin{figure}[!htb]
\centering
\includegraphics[angle=0,scale=0.45,clip]{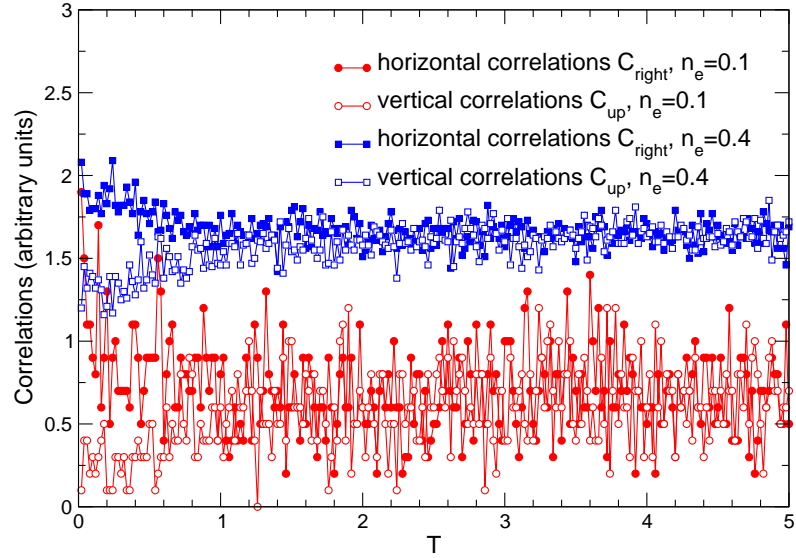}
\caption{\label{fig9}
Horizontal and vertical correlation funtions at densities $n_e=0.1$ and $n_e=0.4$. See text
for the definition of the correlators.}
\end{figure}
%
The algorithm presented in this section is performed at fixed particle density. Each move is
described by the following procedure: Each site is occupied by a spin or hole,
and at every update a site is randomly chosen. If the site is occupied by a spin, its 
direction is reversed and the new configuration is updated depending on the acceptation rate. 
If the site is occupied by a hole, it is exchanged randomly with one of the four
neighboring spins (or possibly holes), mimicking the kinetic transfer but without energy scale $t$. 
The simulations consists in $10000\times N$ updates or $10000$ MC steps at different temperatures
and concentrations of particles. In figure \ref{fig8} are plotted
two final configurations at $n_e=0.4$ and for two different temperatures. At low temperature, the 
holes tend to align along the ferromagnetic couplings and they form an alternating structure of stripes.
For higher temperatures, this alignment breaks into smaller linear hole structures. The system
seems to minimize the frustration by forming linear arrangements along the ferromagnetic couplings.
We should also notice that positioning the holes along the antiferromagnetic couplings should also be a solution as this should lower the energy as well, and this solution appears indeed to depend on the initial conditions for the set of self-consistent equations.
In order to probe the stripe formation and the coherence of these structures, we have plotted 
in figure \ref{fig9} horizontal and vertical correlators $C_{right}$ and $C_{up}$ which are defined as follow: We count the number of sets of two fermions that are aligned horizontally or vertically respectively, which provides an insight of the stripe structure.
We observe in this figure that indeed the horizontal correlations are larger than the vertical ones for temperatures approximately below the coupling value $J$, which is the threshold for the formation of stripes. We also note that, for low densities, these observables are very fluctuating as the probability to find two fermions close to each other is smaller than for denser configurations.
%
\section{Conclusion \label{sect_conc}}
%
Properties of itinerant fermions in a frustrated dilute spin medium with indirect interaction induced by the doping were analyzed in order to probe the effect of spin fluctuations and frustration.
The fermions are found to organize themselves into lines or stripes along the antiferromagnetic or ferromagnetic lines in order to reduce the frustration. These lines arise because of the geometrical constraint imposed by the PUD couplings which are the lines along which frustration can be lowered by distribute the holes. In the ZZD model, these holes would self-organize differently. The effective mass in the mean-field approximation decreases in the regime of low temperature and low density of fermions. This is
consistent with the limiting value of $C_v/T$ in the canonical ensemble at low temperature. In the high
temperature regime, the effective mass is less than unity, contrary to the non-frustrated Ising case.
This suggests that spin fluctuations decrease the effective mass when the system possesses some degree of frustration. The presence of low temperature anomalies at low density in the grand canonical specific heat could be explained by a possible spin-glass solution of the ground state which is not reported by our approximation, or an artifact of this approximation. Such anomalies were however previously observed in the specific heat of spin-ice materials \cite{Pomaranski:2013}, which could point to a complex ground state.
Our model can be generalized with negative couplings different from $-J$, which induces either a ferromagnetic or antiferromagnetic phase transition. It is worth considering in this case the influence of the proximity of a phase transition on the effective mass at low temperature.
A more precise theoretical analysis would effectively lead to more insight, using either diagrammatic methods or dynamical mean field theory. The latter method may be implemented starting from the original Hamiltonian which appears to be similar to a model of spinless fermions with repulsive or attractive next-nearest interactions.

\section*{Acknowledgments}
The authors would like to thank Christophe Chatelain for useful discussions.

\section*{References}
\bibliography{biblio}

\end{document}